%% file: preprint.tex
\documentclass[a4paper,11pt]{article}
\usepackage{graphicx}
\usepackage[T1]{fontenc}
\usepackage{hyperref}
\usepackage{comment}
\usepackage{svg}
\usepackage{multirow}
\usepackage{makecell}
\usepackage{svg}
\usepackage{cite}
\usepackage{balance}

\title{A Systematic Mapping Study on Teaching of Security Concepts in Programming Courses}
\author{
Alina Torbunova \\
Åbo Akademi University \\
Turku, Finland 
\and
Adnan Ashraf \\
Åbo Akademi University \\
Turku, Finland
\and
Ivan Porres \\
Åbo Akademi University \\
Turku, Finland 
}
\date{}

\begin{document}

\maketitle

\begin{abstract}
\textbf{\textit{Context:}} To effectively defend against ever-evolving cybersecurity threats, software systems should be made as secure as possible. To achieve this, software developers should understand potential vulnerabilities and apply secure coding practices. To prepare these skilled professionals, it is important that cybersecurity concepts are included in programming courses taught at universities.

\noindent\textbf{\textit{Objective:}} To present a comprehensive and unbiased literature review on teaching of cybersecurity concepts in programming courses taught at universities. 

\noindent\textbf{\textit{Method:}} We perform a Systematic Mapping Study. We present six research questions, define our selection criteria, and develop a classification scheme.

\noindent\textbf{\textit{Results and Conclusions:}} We select 24 publications. Our results show a wide range of research contributions. We also outline guidelines and identify opportunities for future studies. The guidelines include coverage of security knowledge categories and evaluation of contributions. We suggest that future studies should cover security issues, negative impacts, and countermeasures, as well as apply evaluation techniques that examine students' knowledge. The opportunities for future studies are related to advanced courses, security knowledge frameworks, and programming environments. Furthermore, there is a need of a holistic security framework that covers the security concepts identified in this study and is suitable for education. 

Keywords\textemdash systematic mapping study; cybersecurity; secure programming; teaching; university
\end{abstract}

\input{main}

\bibliographystyle{plain}
\balance
\bibliography{cites.bib}

\end{document}

%% file: main.tex
\section{Introduction}

Along with progress in technology, there is a wide variety of cyberattacks which have become more powerful over time. According to the latest ENISA Threat Landscape report \cite{ENISA_threat_lanskape}, the number and diversity of attacks increased compared to previous years. To reduce security issues in software systems and thus defend them from possible threats, software should be designed with security in mind from the ground up \cite{YuJBY2011} and, in addition, security should be considered in each phase of the software development life cycle \cite{EarwoodYK2021}. 

Developers should create secure software that is resistant to possible attacks \cite{AlmansooriLFSAC2023}. Thus, security awareness is an important skill for any developer \cite{AlmansooriLFSAC2023}. In particular, developers should have knowledge of common vulnerabilities and be able to apply secure coding practices to avoid these \cite{ChiJB2013, TabassumWCL2018}. To teach these skills to future professionals, security should be integrated into the higher education curriculum \cite{SharPSW2022, TabassumWCL2018, WilliamsYYB2014}. 

For researchers, it is important to study the current state of the art to identify areas where further investigation is needed. There is some existing research work on publications related to cybersecurity education~\cite{SvabenskyVC2020}. However, to the best of our knowledge, there are no existing systematic studies on teaching of cybersecurity concepts in programming courses. Therefore, the goal of this Systematic Mapping Study (SMS) is to review research publications that cover teaching of cybersecurity concepts in programming courses in higher education. We answer the following research questions:

\noindent \textbf{RQ1:} Which security concepts are covered?

\noindent \textbf{RQ2:} At which study program and course is the research conducted?\footnote{By study program we mean the level of an academic degree and the study field. Similarly, by course we mean one specific subject in a curriculum.}

\noindent \textbf{RQ3:} What are the contributions of the studies?

\noindent \textbf{RQ4:} Which theoretical frameworks are used?

\noindent \textbf{RQ5:} Which programming environments are used?

\noindent \textbf{RQ6:} How are the contributions evaluated?


This article is structured as follows. In Section \ref{sec-study-design}, we describe how the study was designed and scheduled. We present the results in Section \ref{sec-results}. Then, we provide an evaluation of the validity of this study in Section \ref{sec-validity-evaluation}. Finally, we present and discuss our conclusions in Section \ref{sec-discussion-and-conclusions}.


\section{Study design}
\label{sec-study-design}

To design our SMS, we used the systematic literature reviews and systematic mapping studies guidelines presented in \cite{KitchenhamC2007} and \cite{PetersenVK2015}.  

\subsection{Search strategy}
\label{sec-search-strategy}

In this section, we present chosen search terms, databases, and search strings. 

\subsubsection{Search Terms}

We applied the Population Intervention Comparison Outcome Context (PICOC) criteria \cite{PetticrewR2006} to identify keywords and create search strings. In terms of PICOC, we defined \textit{teaching of programming in higher education} as Population, \textit{cybersecurity concepts} as Intervention, \textit{classification to provide an overview of existing approaches for teaching of cybersecurity concepts at programming courses in higher education} as Outcome, and \textit{academic peer-reviewed publication in the relevant field} as Context. Comparison is not applicable to our study. The final search string is presented in Subsection \ref{subsec-search-strings}.  

\subsubsection{Databases}

We performed the search in the following Digital Libraries (DLs): Institute of Electrical and Electronics Engineers (IEEE) Xplore, Association for Computing Machinery (ACM) DL, ScienceDirect, Scopus, Wiley Online Library,  and Web of Science. SpringerLink was excluded since it was impossible to search only in the selected metadata fields, and the number of returned results was too high.


\subsubsection{Search Strings}
\label{subsec-search-strings}

Table \ref{table:search_string} shows the general structure of the search string. We included in the search the publication title, abstract, and indexing terms or keywords. The structure of the search string was adjusted for each digital library.  For ScienceDirect and Wiley DL\footnote{For Wiley DL, only the search in abstract was considered, since neither the title nor keywords provided meaningful results.}, this led to three separate queries.  Based on the filters offered by DLs, we immediately limited the returned search results. The applied filters included source or document type, subject area, and language (these are also presented in Subsection \ref{sec-inclusion-exclusion-criteria}).


\begin{table}[ht!]
\caption{Search string}
\label{table:search_string}
\centering
\begin{tabular}{|l|}
 \hline
\makecell[l]{(teach OR teaching OR education) AND ("programming" OR "coding") \\AND (secure OR security  OR "cyber security" OR "cybersecurity") \\AND (university OR "higher education")}\\
\hline
\end{tabular}
\end{table}


\subsection{Inclusion and exclusion criteria}
\label{sec-inclusion-exclusion-criteria}
The inclusion criteria is as follows. 1) The study belongs to higher education \textit{AND}
2) The study reports on an approach that aims to apply and improve the teaching of cybersecurity concepts in the scope of programming OR evaluates current state and defines knowledge gaps \textit{AND}
3) The study is written in English \textit{AND} 4) The study is published in a peer-reviewed journal, conference, or workshop.

Additionally, if several papers present the same approach, we included only the most
recent one (applied during full-text screening). The exclusion criteria is opposite to the inclusion criteria. During the title and abstract screening, we applied criterion number 1) only if it was evident that the study does not belong to higher education. 

\subsection{Study selection procedure}

The study selection process included the following steps: title and abstract level screening, full-text level screening, and backward as well as forward snowball sampling. First, we removed duplicates. Second, during the screening at the title and abstract level, we examined the titles and abstracts of the selected publications and applied the inclusion and exclusion criteria. Third, we screened the full texts of the studies selected in the previous step and applied the inclusion and exclusion criteria. Finally, we used the publications selected after full-text level screening for backward and forward snowball sampling. To avoid bias, two researchers (Alina Torbunova and Adnan Ashraf) performed screening and snowball sampling. The results were compared, and disagreements were resolved at meetings.

\subsection{Quality Assessment Checklist and Procedure}
\label{quality-assessment}

The set of selected studies was assessed for their quality. We applied the following checklist for quality assessment. 1) Is teaching of cybersecurity concepts in the scope of programming the main focus of the publication? 2) Does the publication contribute to the improvement of teaching of security concepts in the scope of programming? 3) Is the proposal clearly described? 4) Is the proposal evaluated? 5) Are cybersecurity concepts clearly stated? 6) Is the study already performed?

Two researchers (Alina Torbunova and Adnan Ashraf) performed this task. The results were compared and disagreements were resolved at meetings. To be selected, a study had to satisfy each question in the checklist, as these are crucial for the scope of this SMS and for data extraction.

\subsection{Data Extraction Strategy}


For each of the selected studies, we extracted the title, names of the authors, the year, the type, and the venue. Then, we extracted data items related to each of the six research questions. A publication may contain several values for each research question. To extract the data, we applied the requirement that it should be explicitly specified in the study that the data values in question are related to the contribution of the study.

\subsection{Schedule}
This SMS was performed during December 2023 -- March 2024. The primary studies found by the search string were extracted on 4 December 2023. Backward and forward snowball sampling was performed during January 15--19, 2024.

\section{Results}
\label{sec-results}

The number of publications selected at each stage of the study is presented in Table \ref{publications-at-different-stages}. The initial search in DLs provided 642 results. We selected 24 publications after applying inclusion and exclusion criteria, snowball sampling, and quality assessment.




\begin{table}[ht!]
\caption{Number of publications at each stage of study}
\label{publications-at-different-stages}
\centering
\begin{tabular}{|l|c|}
 \hline
 Stage  & Nr. of publications \\  
 \hline
Search string & 642\\
 \hline
After removing duplicated & 484\\
 \hline
After title \& abstract screening & 85 \\
 \hline
After full text screening & 24 \\
 \hline
After backward snowball sampling & 30 \\
\hline
After forward snowball sampling & 44 \\
\hline
After quality assessment & 24 \\
 \hline
\end{tabular} 
\end{table}

Table \ref{table:publications-list} presents the list of publications selected for this study. The table includes Study IDs (SIDs) given to the selected studies, along with a reference to the source, the type of venue\footnote{Among venue types, a symposium was classified as a conference.}, and the year of publication. These studies cover the period from 2010 to 2023. The number of studies in each year ranges from zero to three, with an increase in 2019 with six studies published in that year. 

\begin{table}[ht!]
\caption{Selected publications}
\label{table:publications-list}
\centering
\begin{tabular}{|l|l|l|l||l|l|l|l|}
 \hline
SID & Ref. & Venue & Year & SID & Ref. & Venue & Year\\  \hline
S1 & \cite{AbernathyYHXBW2017} & Conf. & 2017 & S13 & \cite{MdunyelwaFN2019} & Conf. & 2019\\  \hline
S2 & \cite{AlmansooriLFMSAC2020} & Conf. & 2020 & S14 & \cite{Pawelczak2020} & Conf. & 2020\\  \hline
S3 & \cite{AlmansooriLFSAC2023} & Conf. & 2023 & S15 & \cite{SharPSW2022} & Conf. & 2022\\  \hline
S4 & \cite{AzizSH2016} & Conf. & 2016 & S16 & \cite{TabassumWCL2018} & Conf. & 2018\\  \hline
S5 & \cite{BandiFB2019} & Journal & 2019 & S17 & \cite{TangD2019} & Conf. & 2019\\  \hline
S6 & \cite{BishopDFNNBZ2019} & Conf. & 2019 & S18 & \cite{TaylorK2016} & Journal & 2016\\  \hline
S7 & \cite{BishopDDNNZ2017} & Conf. & 2017 & S19 & \cite{WenK2019} & Conf. & 2019\\  \hline
S8 & \cite{DossYCAF2019} & Journal & 2019 & S20 & \cite{WhitneyRCZ2015} & Conf. & 2015\\  \hline
S9 & \cite{EarwoodYK2021} & Conf. & 2021 & S21 & \cite{WilliamsYYB2014} & Journal & 2014\\  \hline
S10 & \cite{Elva2022} & Conf. & 2022 & S22 & \cite{YuJBY2011} & Conf. & 2011\\  \hline
S11 & \cite{GhiglieriS2016} & Conf. & 2016 & S23 & \cite{PeltsvergerK2010} & Conf. & 2010\\  \hline
S12 & \cite{Haywood2013} & Conf. & 2013 & S24 & \cite{ChiJB2013} & Conf. & 2013\\  \hline
\end{tabular} 
\end{table}

The categorization of data extracted for each of the research questions is presented in Sections \ref{subsec-security-consepts} to \ref{subsec-evaluation approach}. In S16 we consider only the data related to the clinic approach, since the tool ESIDE which is compared to the clinic is already presented in detail in S20 (see Subsection \ref{subsec-proposed-approach} for more information about contributions).

\subsection{Security Concepts (RQ1)}
\label{subsec-security-consepts}
For the first research question, we divided the topics related to security concepts into four main categories: \textit{General security knowledge}, \textit{Security issues}, \textit{Negative impacts}, and \textit{Countermeasures}. The general idea of these categories is inspired by \cite{Elva2022} and \cite{WenK2019}, but we selected different naming conventions and have an additional category related to general knowledge. We present each category of security concepts in detail in the following subsections. Then, we analyze how a combination of security issues, negative impacts, and countermeasures is covered in studies. General security knowledge is not included in this analysis, since this category includes broad concepts.


\subsubsection{General security knowledge}
\label{subsubsec-general-security-knowledge}

The General security knowledge category includes general concepts. These are mentioned as awareness of unsafe functions (S3), security topics (S3), security engineering (S11 and S22), threat modeling (S6, S22), robustness (S6), security vulnerabilities (S10 and S22), risks (S10 and S18), threats (S10 and S22), attacks (S10), trust (S10), trustworthiness (S10), balancing security properties (S10), authenticity (S22), accountability (S22), assets (S22), poor programming practices and possible impacts of these (S22), security violations (S22), as well as confidentiality, integrity, and availability (S22). In some studies, theoretical frameworks (see Table \ref{table:theoretical-frameworks}) were presented in a general way as a part of the contribution (S4, S8, S11, and S15).


\subsubsection{Security issues}
\label{subsubsec-security-issues}
Topics related to security issues or, in other words, vulnerabilities, are presented in Table \ref{table:security-issues}. These topics are divided into two main categories: \textit{General} and \textit{Language-based}. The idea of the latter category is inspired by \cite{BandiFB2019}. The choice of categories is based on the extracted data as well as on the own understanding of the authors of this study, since there was a different theoretical base utilized by researchers and thus affected terminology used in the selected studies. In general security issues, the category of file system vulnerabilities includes file input and output validation vulnerabilities (S4) as well as storing sensitive data on external storage (S8). Among general issues, the majority of studies refer to insecure communication, while improper authentication and session management as well as side-channel vulnerability are mentioned in one study each. 

In language-based security issues in Table \ref{table:security-issues}, there are three categories that include concrete subtopics. In memory management vulnerabilities, these are buffer overflow (S2--S6, S11, S14, S18, S21, and S24), integer overflow (S2, S5, S10, S18, S21, and S22), integer underflow (S18), memory access vulnerabilities (S10), and concurrency vulnerabilities (S4). Injection flaws include OS command injection (S2), SQL injection (S4, S11, S16, S20, S23, and S24), cross-site scripting (S4, S11, S15, S16, S20, and S24), and image tag vulnerability (S24). Finally, output vulnerabilities include format string vulnerability (S2, S14), output encoding vulnerability (S4, S16, S19, and S20), and output escaping vulnerability (S4 and S19). Among language-based security issues, the majority of studies refer to memory management vulnerabilities, while cross-site request forgery vulnerability is the least covered topic.


\begin{table}[ht!]
\caption{Security issues}
\label{table:security-issues}
\centering
\begin{tabular}{|l|l|l|} 
\hline
Category & Subcategory & SID \\
\hline
\multirow{7}{*}{General} 
& Improper access control & S1, S4, S8, S16 \\\cline{2-3}
& \makecell[l]{Improper authentication\\and session management} & S4 \\\cline{2-3}
& Insecure communication & S1, S4, S5, S8, S11 \\\cline{2-3}
& Improper configuration & S1, S4, S8 \\\cline{2-3}
& Weak cryptography & S4, S5, S11 \\\cline{2-3}
& Side-channel vulnerability & S17 \\\cline{2-3}
& File system vulnerabilities & S4, S8 \\\cline{2-3}
\hline
\multirow{8}{*}{\makecell[l]{Language-\\based}}
& Improper input validation & \makecell[l]{S1, S5, S10, S14, S16,\\S18--S20, S22} \\\cline{2-3}
& \makecell[l]{Memory management\\vulnerabilities} & \makecell[l]{S2--S6, S10, S11, S14, S18,\\S21, S22, S24} \\\cline{2-3}
& Injection flaws & \makecell[l]{S2, S4, S11, S15, S16, S20,\\S23, S24} \\\cline{2-3}
& Output vulnerabilities & S2, S4, S14, S16, S19, S20 \\\cline{2-3}
& Improper error handling & S4, S5, S9, S10 \\\cline{2-3}
& Improper logging & S1, S8, S10 \\\cline{2-3}
& Cross-site request forgery & S4, S11 \\\cline{2-3}
& \makecell[l]{Insecure general coding\\practices} & S4, S5, S9, S14, S21 \\\cline{2-3}
\hline
\end{tabular}
\end{table}


\subsubsection{Negative impacts}
\label{subsubsec-negative-impacts}
Negative impacts are presented in Table \ref{table:negative-impacts}. These topics are divided into three main categories: \textit{Attacks}, \textit{Unexpected behavior}, and \textit{Security violations}. Studies included in the first eight attack categories provide an exact name of the attack that is a part of the contribution. The last category of attacks, other malware attacks, includes studies in which the exact attack names were not provided. Injection attacks are presented with the following three types of attack: SQL injection (S4, S11, S16, S20, and S23), HTML injection (S4), and cross-site scripting (S4, S11, S15, S16, S19, and S20). The general category of unexpected behavior includes studies (S10 and S21) in which a more concrete example was not provided. In S4, S8, S11, and S21 negative impacts are taught in practice, for example, students need to perform an attack to gain unauthorized access to the system. Most of the studies in the Attacks category are on the injection attacks. Similarly, code crash is the leading subcategory under Unexpected behaviors. Unauthorized access is the leading one under Security violations. 

\begin{table}[ht!]
\caption{Negative impacts}
\label{table:negative-impacts}
\centering
\begin{tabular}{|l|l|l|} 
\hline
Category & Subcategory & SID \\
\hline
\multirow{8}{*}{Attacks} 
& Parameter tampering & S4 \\\cline{2-3}
& System hijacking & S10 \\\cline{2-3}
& Malicious execution & S4 \\\cline{2-3}
& Injection & \makecell[l]{S4, S11, S15, S16,\\S19, S20, S23} \\\cline{2-3}
& Denial-of-service & S4, S10, S16 \\\cline{2-3}
& Eavesdropping & S8, S11 \\\cline{2-3}
& Man-in-the-middle & S9, S16 \\\cline{2-3}
& Side-channel attacks & S17 \\\cline{2-3}
& Other malware attacks & \makecell[l]{S8, S11, S21, S22} \\\cline{2-3}
\hline
\multirow{5}{*}{\makecell[l]{Unexpected\\behavior}}
& Code crash & S10, S21, S22 \\\cline{2-3}
& Infinite loop & S10 \\\cline{2-3}
& Memory corruption & S10, S14 \\\cline{2-3}
& \makecell[l]{Function parameter\\modification} & S14 \\\cline{2-3}
& General & S10, S21 \\\cline{2-3}
\hline
\multirow{2}{*}{Security violations}
& Unauthorized access & \makecell[l]{S7, S8, S10, S11,\\S17} \\\cline{2-3}
& Unauthorized modification & S7, S8, S10 \\\cline{2-3}
\hline
\end{tabular}
\end{table}



\subsubsection{Countermeasures}
\label{subsubsec-countermeasures}

Countermeasures are presented in Table \ref{table:countermeasures}. These topics are divided into two main categories: \textit{Secure coding practices} and \textit{Other approaches} and define practices that contribute to a more secure code. As a framework for the category of secure coding practices, we utilized OWASP Secure Coding Practices\footnote{The OWASP category session management is not present, since no topics could be mapped to this category.} \cite{OWASPSecCod}, while other approaches are based only on our categorization of the extracted data. We defined subtopics for three categories of secure coding practices. In authentication and password management, these are authentication (S11--S13, S15, and S17) and password-related practices (S7, S8, S12, S16, and S17). Cryptographic practices are covered with five concrete topics: encryption and decryption (S5, S7--S9, S11--S13, and S16), hashing (S5, S11, S12, S16, and S17), secure random number generation or salting (S5, S7, S9, and S16), cryptographic protocols (S12 and S17), and message authentication (S5 and S12). Finally, in the error handling and logging category there are two subcategories: error handling (S5, S7, S9, S10, S12, and S21) and logging (S1, S8, S10, and S17). 

Among other approaches in Table \ref{table:countermeasures}, the code review category includes studies that utilize static code analysis tools (S1, S2, S8, S15, S20, and S24), compiler messages (S3, S7, and S14), and manual code review (S6, S10, S16, and S22) as ways to improve security of the code. By documentation resources (S2 and S3) we mean library documentation and other information resources that are utilized by students. Input validation is the leading topic in secure coding practices, while code review is the leading one among other approaches.

\begin{table}[ht!]
\caption{Countermeasures}
\label{table:countermeasures}
\centering
\begin{tabular}{|l|l|l|} 
\hline
Category & Subcategory & SID \\
\hline
\multirow{13}{*}{\makecell[l]{Secure\\coding\\practices}}
& Input validation & \makecell[l]{S1--S7, S10, S13--S22, S24}\\\cline{2-3}
& Output encoding & \makecell[l]{S2, S4, S7, S14, S16, S19,\\S20, S22} \\\cline{2-3}
& \makecell[l]{Authentication and\\password management} & \makecell[l]{S7, S8, S11--S13, S15--S17} \\\cline{2-3}
& Access control & \makecell[l]{S1, S7, S8, S11--S13, S16,\\S17} \\\cline{2-3}
& Cryptographic practices & \makecell[l]{S5, S7--S9, S11--S13, S16,\\S17} \\\cline{2-3}
& Error handling and logging & \makecell[l]{S1, S5, S7--S10, S12, S17,\\S21} \\\cline{2-3}
& Data protection & \makecell[l]{S1, S5, S7--S13,\\S17} \\\cline{2-3}
& Communication security & S1, S5, S8, S11, S12 \\\cline{2-3}
& System configuration & \makecell[l]{S1, S7, S8, S13, S15, S17} \\\cline{2-3}
& Database security & S8, S13, S16, S23 \\\cline{2-3}
& File management & S1, S7--S9, S17 \\\cline{2-3}
& Memory management & \makecell[l]{S2--S7, S10, S14, S17, S18,\\S21, S22} \\\cline{2-3}
& General coding practices & \makecell[l]{S2, S3, S5, S7, S9, S10,\\S12, S14, S21} \\\cline{2-3}
\hline
\multirow{2}{*}{\makecell[l]{Other}}
& Code review & \makecell[l]{S1--S3, S6--S8, S10,\\S14--S16, S20, S22, S24} \\\cline{2-3}
& Documentation resources & S2, S3 \\\cline{2-3}
\hline
\end{tabular}
\end{table}



\subsubsection{Coverage of Security Issues, Negative Impacts, and Countermeasures}
\label{subsubsec-sec-issues-neg-impacts-countermeasures}

\begin{figure}[ht!]
   \centering
   \includegraphics[width=7.7cm]{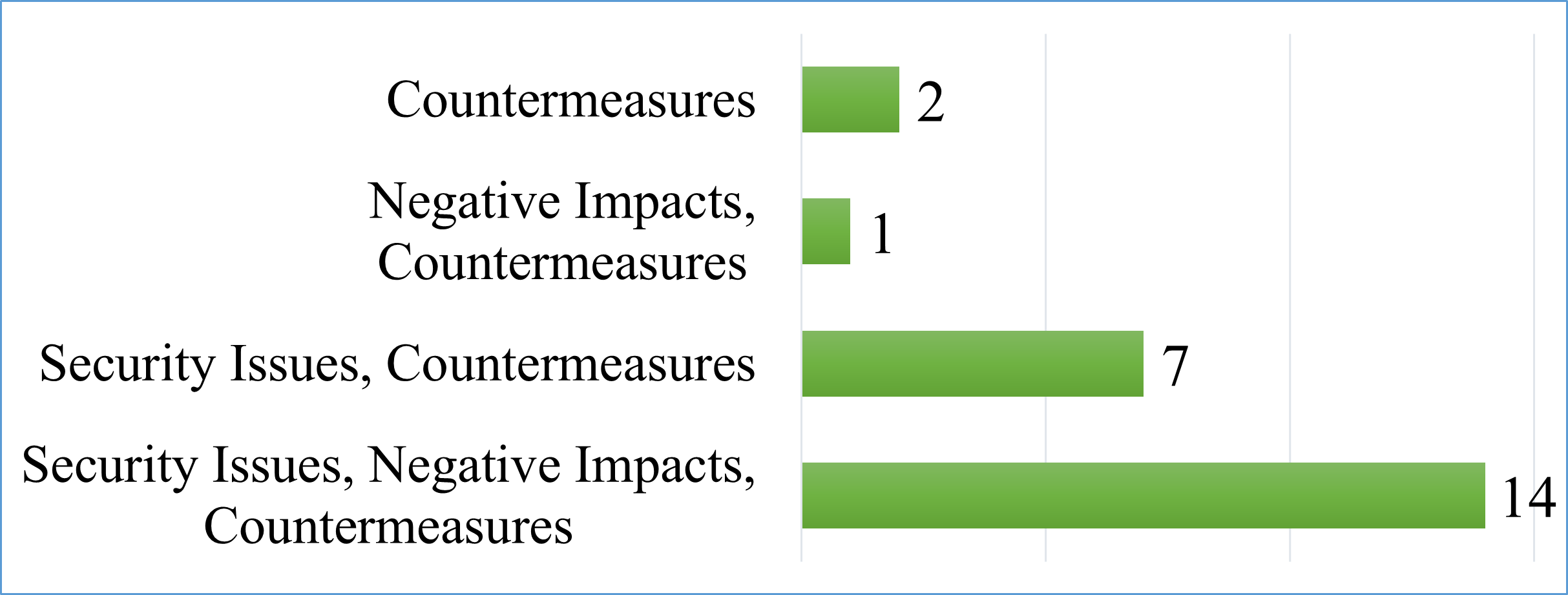}
    \caption{Security knowledge categories covered in studies}
    \label{fig:categories}
\end{figure}

We analyzed how a combination of security issues, negative impacts, and countermeasures is covered in studies. Figure \ref{fig:categories} shows how these are distributed. In 14 studies, all three categories are covered. The combination of negative impacts and countermeasures is covered in one study, while the combination of security issues and countermeasures is covered in seven studies. Countermeasures are only covered in two studies. We observed that a general trend is to combine security issues, negative impacts, and countermeasures, but there are studies that miss at least one security knowledge category. As a guideline, we concluded that studies should cover all three knowledge categories.

\subsection{Study Programs and Courses (RQ2)}
\label{subsec-program-and-course}

Table \ref{table:study-programmes-list} presents types and names of study programs where the research was conducted. We identified \textit{undergraduate} (includes bachelor and diploma degrees), \textit{graduate} (includes master's and doctoral degrees), and \textit{unspecified} (if no information could be derived), while names of these programs were categorized as \textit{Computer Science (CS)}, \textit{non-Computer Science (non-CS)}, and \textit{unspecified}. Some studies belong to several categories if courses are offered to students at different study programs. The majority of courses belong to undergraduate CS study programs.

\begin{table}[ht!]
\caption{Study programs}
\label{table:study-programmes-list}
\centering
\begin{tabular}{|l|l|l|} 
\hline
Level & Name & Study ID \\
\hline
\multirow{2}{*}{Undergraduate} 
& Computer Science & \makecell[l]{S2--S5, S9, S10, S12,\\S16, S18, S20--S23} \\\cline{2-3}
& Non-Computer Science & \makecell[l]{S4, S5, S13, S14, S15,\\S19, S24} \\\cline{2-3}
\hline
\multirow{2}{*}{Graduate} & Computer Science & S16, S17, S20 \\\cline{2-3}
& Unspecified & S8 \\
\hline
\multirow{2}{*}{Unspecified} & Computer Science & S1, S6 \\\cline{2-3}
& Unspecified & S7, S8, S11 \\ 
\hline
\end{tabular}
\end{table}

Course levels and their names are presented in Table \ref{table:courses-list}. We categorized courses as \textit{introductory}, \textit{intermediate}, \textit{advanced}, and \textit{unspecified}. The majority of courses are introductory. We performed an additional grouping for course names. The category \textit{Programming} in introductory, intermediate, and courses of unspecified level includes courses with different original names. Since some studies evaluate the contribution at several courses, there are publications that belong to different categories. The majority of courses are introductory, followed by the intermediate level.

Programming is usually taught at undergraduate programs, but some studies cover graduate programs and thus advanced courses. For advanced courses that require programming, we identified a research opportunity to study knowledge gaps as well as how security concepts can be integrated into these courses. 

\begin{table}[ht!]
\caption{Courses}
\label{table:courses-list}
\centering
\begin{tabular}{|l|l|l|} 
\hline
Level & Name & Study ID \\
\hline
\multirow{7}{*}{Introductory} & Programming & \makecell[l]{S5, S6, S9,\\S12, S14,\\S21, S22} \\\cline{2-3}
& Computer Systems & S2, S3 \\\cline{2-3}
& IT Security & S11 \\\cline{2-3}
& Computer Science 0, 1, and 2 & S18 \\\cline{2-3}
& Database Systems & S23 \\
\hline
\multirow{5}{*}{Intermediate} & Programming & S22, S24 \\\cline{2-3}
& Secure Software Engineering & S10 \\\cline{2-3}
& Web Application Security & S13 \\\cline{2-3}
& IT Security & S4 \\\cline{2-3}
& Web Application Development 2 & S15 \\\cline{2-3}
& Network-Based Multimedia & S24 \\
\hline
\multirow{3}{*}{Advanced} & Secure Software Engineering & S8\\\cline{2-3}
& Network Based Application Development & S16, S20 \\\cline{2-3}
& \makecell[l]{Internet Security, Information Security and\\Privacy} & S17 \\
\hline
\multirow{5}{*}{Unspecified} & Programming & \makecell[l]{S1, S7, S8,\\S24}\\\cline{2-3}
& \makecell[l]{Networking, Operating Systems} & S6 \\\cline{2-3}
& Computer Security & S6, S7 \\\cline{2-3}
& Web Design & S24 \\\cline{2-3}
& Software Security & S19 \\\cline{2-3}
& Unspecified & S7 \\\cline{2-3}
\hline
\end{tabular}
\end{table}

\subsection{Contributions (RQ3)}
\label{subsec-proposed-approach}


Table \ref{table:proposed-approaches} shows the contributions of the selected studies. These are divided into eight categories. \textit{Course modules} include teaching of security concepts in parallel with the course material (S5 and S14), separate lessons related to security concepts (S9, S10, S12, S13, S15, S21, S22, and S24), which can also be laboratory-based (S18), as well as integration of security concepts to course assignments, projects, and/or quizes (S9, S13, S15, S18, and S24). \textit{E-learning platforms} include theoretical information (S4, S19, and S24), practical tasks with the use of additional tools such as SCALT (S4), WebGoat (S4), FindBugs (S24), and Splint (S24), quizzes (S24), as well as tools for submission and classification of assignments (ChainGrader in S17). \textit{Labs} include both theoretical components and hands-on exercises (S8 and S17), as well as serve as the supplementary approach in addition to teaching of security concepts along with the course material (S5). 

Three studies identify \textit{knowledge gaps} related to memory management vulnerabilities, injection flaws and attacks, input validation, output encoding, database security, general coding practices, code review, and usage of documentation resources. These gaps are explored by analyzing students' course assignments (S2 and S23), course books, code provided by instructors, and other course materials (S2), as well as by organizing coding interviews with students (S3). \textit{Static code analysis tools} proposed by researchers include SACH (S1) and ESIDE (S20). These tools find vulnerabilities in the code and provide theoretical information as well as hints to fix these issues. The \textit{clinic} approach (S6 and S16) means that students can visit separate in person sessions where they show their code and receive feedback. \textit{Teaching approaches} include TRAC (S10) and the contextualized learning approach (S19). Finally, the proposed \textit{taxonomies} aim to evaluate students' understanding of security concepts (S3) or consist of the concept map for knowledge assessment (S7) which is further applied to develop a set of questions. Some studies include more than one contribution, and thus belong to several categories. 

The majority of studies propose a course module. E-learning platform is the second most popular contribution, while in the rest of categories the number of studies is more evenly distributed. Selected studies show a variety of contributions, and these are closely connected with the learning objectives of the courses.


\begin{table}[ht!]
\caption{Contributions}
\label{table:proposed-approaches}
\centering
\begin{tabular}{|l|l|}
 \hline
 Contribution & Study IDs\\  
 \hline
 Course Modules & \makecell[l]{S5, S9, S10, S12, S13--S15, S18, S21,\\S22, S24}\\
 \hline
 E-learning Platforms & S4, S11, S17, S19, S24\\
 \hline
 Labs & S5, S8, S17 \\
 \hline
 Knowledge gaps & S2, S3, S23\\
 \hline
 Code Analysis Tools &  S1, S20\\
 \hline
 Clinics & S6, S16\\
 \hline
 Teaching Approaches & S10, S19\\
 \hline
 Taxonomies & S3, S7\\
 \hline
\end{tabular} 
\end{table}

\subsection{Utilized theoretical frameworks (RQ4))}
\label{subsec-theoretical-frameworks}

Table \ref{table:theoretical-frameworks} presents the theoretical frameworks that were used in the selected studies. We divided these frameworks into three main groups: \textit{security}-, \textit{learning}-related and \textit{unspecified}. Some publications refer to more than one framework. There are frameworks that are updated at certain time intervals or include additional subcategories, and different versions of the frameworks were utilized by researchers. We did not consider these differences in this study. 

Among security-related frameworks, the most popular is OWASP Top 10, followed by ACM curriculum guidelines. The category of security books and publications includes \cite{ViegaM2001, HowardL2003, HowardLV2005, TsipenyukCM2005} from S18, \cite{StallingsB2007} and the NIST Computer Security Handbook from S22, as well as \cite{Cenzic2009} from S23. In learning-related frameworks, each category contains one study. We merged some categories if they belonged to the same study. Finally, the unspecified category includes studies in which we could not extract any concrete or relevant information. 

We observed that there is a wide range of security frameworks that researchers refer to. Some frameworks are focused only on one category of security concepts, for example, secure coding standards or security issues. In addition, among the frameworks that cover curricula recommendations or work skill requirements, we found that they lack some topics covered in the selected studies. For example, ACM curriculum guidelines are missing topics such as format string vulnerability, output encoding vulnerability, side-channel vulnerability, and output encoding as a countermeasure. Thus, we conclude that research related to a holistic framework that is suitable for learning security concepts at programming courses is needed.

\begin{table}[ht!]
\caption{Applied theoretical frameworks}
\label{table:theoretical-frameworks}
\centering
\begin{tabular}{|l|l|}
\hline
Category & Framework (Study IDs)\\  
\hline
\multirow{11}{*}{Security} & OWASP Top 10 \cite{OWASPTop10} (S4, S11, S15, S18, S20)\\\cline{2--2}
& ACM Curriculum Guidelines \cite{ACMCurricula} (S10, S13, S14, S21)\\\cline{2--2}
& \makecell[l]{SEI CERT Secure Coding Standard for Java \cite{SEICERTJava}\\(S1, S4, S8)}\\\cline{2--2}
& Security books and publications (S18, S22, S23)\\\cline{2--2}
& SANS/CWE Top 25 \cite{CWETop25} (S4, S10, S18)\\\cline{2--2}
& SEI CERT C Secure Coding Standard \cite{SEICERTC} (S4, S14)\\\cline{2--2}
& CWE \cite{CWE} (S2, S9)\\\cline{2--2}
& OWASP Secure Coding Practices \cite{OWASPSecCod} (S13)\\\cline{2--2}
& SEI CERT C++ Secure Coding Standard \cite{SEICERTC++} (S4)\\\cline{2--2}
& NICE Cybersecurity Workforce Framework \cite{NICE} (S9)\\\cline{2--2}
& SEI CERT Top 10 Secure Coding Practices \cite{SEICERTTop10} (S18)\\\cline{2--2}
\hline
\multirow{7}{*}{Learning} & SOLO Taxonomy (S3)\\\cline{2--2}
& Brain-Compatible Learning Principles (S6)\\\cline{2--2}
& Theories for Concept Inventory (S7)\\\cline{2--2}
& Model-Eliciting Activity (S9)\\\cline{2--2}
& Threaded teaching approach (S10)\\\cline{2--2}
& \makecell[l]{Bloom's Taxonomy and Principles of Active Learning\\(S18)}\\\cline{2--2}
& \makecell[l]{Context-Based Knowledge and Contextualized Learning\\(S19)}\\ \hline
\multicolumn{1}{|c|}{Unspecified} & (S5, S12, S16, S17, S19, S24)\\
 \hline
\end{tabular} 
\end{table}

\subsection{Programming environments (RQ5))}
\label{subsec-programming-environments}

Table \ref{table:programming-environments} presents the programming environments used in the courses. Among these, the most popular programming language is Java\footnote{In S1 and S8, Java defined as a programming language, but the contribution is specifically for the development of Android applications.}, followed by C and C++. Sometimes a concrete programming language is not specified (S6 and S10), but in one study students could freely choose a compatible programming language while using the .NET framework (S13). In S17, in addition to C and Bash scripting, Intel SGX and Blockchain programming platform are utilized for programming on trusted platforms. Some studies belong to several categories. 

Among programming environments, we observed a wide range of programming languages, but Python is covered in only one study. We conclude that there is a need for research related to security concepts applicable to Python, since it is one of the most commonly taught programming languages, especially in introductory programming courses \cite{TaylorK2016}.


\begin{table}[ht!]
\caption{Programming environments}
\label{table:programming-environments}
\centering
\begin{tabular}{|l|l|}
 \hline
 Programming environment & Study IDs\\  
 \hline
C & S2--S4, S6, S7, S14, S17\\
\hline
C++ & S2, S4, S7, S17, S18, S24\\
\hline
Java & \makecell[l]{S1, S4, S5, S8, S9, S12, S16,\\S18, S20--S22, S24}\\
\hline
JavaScript & S11, S15, S19\\
\hline
SQL & S11, S23\\
\hline
HTML & S15, S19, S24\\
\hline
CSS & S15\\
\hline
Python & S18\\
\hline
PHP & S19, S23\\
\hline
.NET framework & S13\\
\hline
\makecell[l]{Bash scripting, Intel SGX,\\Blockchain programming platform} & S17\\
\hline
Unspecified & S6, S10\\
 \hline
\end{tabular} 
\end{table}

\subsection{Evaluation of contributions (RQ6)}
\label{subsec-evaluation approach}

Table \ref{table:evaluations} shows the approaches applied by researchers to evaluate the contributions of their studies. These are divided into 13 categories and include surveys of several types, analysis of code submitted either as a course assignment or as a part of an exam, interviews, feedback, grades, observations, interaction recordings, and analysis of code materials. Depending on the content of surveys, we defined them as a \textit{survey} (S1, S4--S6, S8, S10--S12, S14--S17, S19--S22, and S24) if there are no theoretical questions included, while if they are present, such approaches are categorized as a \textit{test survey} (S4, S7, S13, S16, and S18--S20) or as a \textit{coding survey} if there are tasks related to coding (S3 and S18). Surveys and test surveys are additionally divided into \textit{pre-} and \textit{post-}, depending on when they are applied to the subject of the study. In addition to coding surveys, coding skills were also evaluated with the help of code analysis that was performed with \textit{code analysis tools} (S2, S15, and S23) or \textit{manually} (S9, S10, S13, and S14) as well as with grades (S1, S6, S8, S11, and S14). Another approach to evaluate a contribution is an \textit{interview} (S3, S15, and S20) that is applied to analyze students' perceptions or practical coding behavior. The \textit{feedback} category (S10 and S17) includes informal comments collected from students. The interactions between the students were analyzed with the help of \textit{observations} (S10), while the interactions with the teaching assistant or the proposed tool are classified as \textit{ interaction recordings} (S16, S17 and S20). \textit{Manual analysis of course materials} (S2) means that researchers analyzed textbooks and lecture materials.

We observed a variety of techniques used to evaluate the research contributions. In the majority of studies, a post-survey is applied, but this technique does not allow researchers to evaluate students' skills. Another interesting observation is that if a pre-survey or a pre-test survey was applied, the corresponding post-approach was always utilized too, but both post-survey and post-test survey could be applied without the pre-approach. In the majority of studies, the main research contribution is evaluated with more than one approach. Among studies that belong to several categories, there are cases when the approach was evaluated with several methods or it could be applied at different courses or evaluated several times (for example, in different years or on different groups of participants). We concluded that all studies should apply test surveys, coding surveys, manual and automated analysis of students' code as well as grades, since these techniques allow researchers to evaluate students' understanding of cybersecurity concepts.

\begin{table}[ht!]
\caption{Evaluation of contributions}
\label{table:evaluations}
\centering
\begin{tabular}{|l|l|}
 \hline
 Evaluation & Study IDs \\  
 \hline
 Pre-surveys   &  \makecell[l]{S5, S6, S12, S20--S22, S24} \\
 \hline
 Post-surveys   &\makecell[l]{S1, S4--S6, S8, S10--S12,\\S14--S17, S19--S22, S24}\\
 \hline
 Pre-test surveys  &  S7, S13, S18, S19, S20\\
 \hline
 Post-test surveys  &  \makecell[l]{S4, S7, S13, S16, S18, S19, S20}\\
 \hline
 Coding surveys  &  S3, S18\\
 \hline
 Code analysis with tools  & S2, S15, S23\\
  \hline
 Manual analysis of code  &  S9, S10, S13, S14\\
 \hline
 Grading  &  S1, S6, S8, S11, S14 \\
 \hline
 Interviews  & S3, S15, S20\\
 \hline
 Feedback  &  S10, S17\\
 \hline
 Observations & S10\\
 \hline
 Interaction recordings  &  S16, S17, S20\\
 \hline
 Manual analysis of course materials  &  S2\\
 \hline
\end{tabular}
\end{table}

\section{Validity evaluation}
\label{sec-validity-evaluation}
To evaluate the validity of our study, we identify threats to the validity of the conclusion, internal, construct and external \cite{WohlinRHORW2012, CookC1979}. First, among threats to \textit{conclusion validity}, we define credibility of research findings. To mitigate this threat, we based our findings on the extracted data and resolved disagreements at meetings. Second, selection of primary studies is a threat to \textit{internal validity}. To mitigate this threat, we followed established guidelines for performing SMS \cite{KitchenhamC2007, PetersenVK2015} and clearly documented the whole process. We used six comprehensive digital libraries to form the initial set of our studies. Then, to avoid bias, two researchers (Alina Torbunova and Adnan Ashraf) independently screened the publications and resolved disagreements at the meetings. Third, a threat to \textit{external validity} is the relevance of the findings to the current research context. To mitigate this threat, we applied search, inclusion, and exclusion criteria as well as quality assessment to exclude irrelevant studies. A threat to \textit{construct validity} is bias in the interpretation of the results based on our own expectations from the experience. This is not applicable to our study, since we did not have predetermined expectations.


\section{Discussion and conclusions}
\label{sec-discussion-and-conclusions}
In this paper, we presented the results of a systematic mapping study of research publications on teaching of cybersecurity concepts in programming courses in higher education. We selected 24 publications which cover the period from 2010 to 2023. We describe guidelines and outline some opportunities for future studies.

\textit{Guidelines}: We outlined guidelines related to the coverage of security knowledge categories and evaluation of contributions. First, we observed that a general trend is to combine security issues, negative impacts, and countermeasures, but there are studies that do not cover all three categories. We suggest that studies should cover all three knowledge categories to encompass the whole spectrum. Second, we noticed that the majority of studies utilized a survey, in particular a post-survey, to evaluate a contribution. This technique does not allow researchers to evaluate students' understanding of cybersecurity concepts, since it does not include theoretical questions or programming tasks. Thus, we recommend that all studies should apply test surveys, coding surveys, manual and automated analysis of students' code as well as grades.

\textit{Opportunities}: We identified research opportunities related to courses, frameworks, and programming environments. The first research opportunity is to study knowledge gaps as well as how security concepts can be integrated into advanced courses that require programming. The second research opportunity is related to a holistic security framework that is suitable for learning security concepts at programming courses. We observed that utilized frameworks either concentrated on one knowledge category or were missing some topics. Among programming environments, we observed that Python is covered by only one study, although this programming language is commonly taught. Thus, the third research opportunity is related to security concepts applicable to Python.

\section*{Acknowledgments}
This work has received funding from the Finnish Ministry of Education and Culture under grant agreement OKM/60/522/2022. Adnan Ashraf received a research grant from the Ulla Tuominen Foundation.

%% file: preprint.bbl
\begin{thebibliography}{10}

\bibitem{AbernathyYHXBW2017}
Aakiel Abernathy, Xiaohong Yuan, Edward Hill, Jinsheng Xu, Kelvin Bryant, and Kenneth Williams.
\newblock {SACH}: A tool for assisting {S}ecure {A}ndroid application development.
\newblock In {\em SoutheastCon 2017}, pages 1--4, 2017.

\bibitem{ACMCurricula}
ACM.
\newblock Curricula recommendations.
\newblock Accessed: 2024-02-15.

\bibitem{AlmansooriLFMSAC2020}
Majed Almansoori, Jessica Lam, Elias Fang, Kieran Mulligan, Adalbert Gerald~Soosai Raj, and Rahul Chatterjee.
\newblock How secure are our computer systems courses?
\newblock In Anthony~V. Robins, Adon Moskal, Amy~J. Ko, and Ren{\'{e}}e McCauley, editors, {\em {ICER} 2020: International Computing Education Research Conference, Virtual Event, New Zealand, August 10-12, 2020}, pages 271--281. {ACM}, 2020.

\bibitem{AlmansooriLFSAC2023}
Majed Almansoori, Jessica Lam, Elias Fang, Adalbert Gerald~Soosai Raj, and Rahul Chatterjee.
\newblock Towards finding the missing pieces to teach secure programming skills to students.
\newblock In Maureen Doyle, Ben Stephenson, Brian Dorn, Leen{-}Kiat Soh, and Lina Battestilli, editors, {\em Proceedings of the 54th {ACM} Technical Symposium on Computer Science Education, Volume 1, {SIGCSE} 2023, Toronto, ON, Canada, March 15-18, 2023}, pages 973--979. {ACM}, 2023.

\bibitem{AzizSH2016}
Normaziah~A. Aziz, Siti Nurul~Zulaiha Shamsuddin, and Nur~Asnida Hassan.
\newblock Inculcating secure coding for beginners.
\newblock In {\em 2016 International Conference on Informatics and Computing (ICIC)}, pages 164--168, 2016.

\bibitem{BandiFB2019}
Ajay Bandi, Abdelaziz Fellah, and Harish Bondalapati.
\newblock Embedding security concepts in introductory programming courses.
\newblock {\em Journal of Computing Sciences in Colleges}, 34(4):78--89, 2019.

\bibitem{BishopDDNNZ2017}
Matt Bishop, Jun Dai, Melissa Dark, Ida Ngambeki, Phillip Nico, and Minghua Zhu.
\newblock Evaluating secure programming knowledge.
\newblock In Matt Bishop, Lynn Futcher, Natalia~G. Miloslavskaya, and Marianthi Theocharidou, editors, {\em Information Security Education for a Global Digital Society - 10th {IFIP} {WG} 11.8 World Conference, {WISE} 10, Rome, Italy, May 29-31, 2017, Proceedings}, volume 503 of {\em {IFIP} Advances in Information and Communication Technology}, pages 51--62. Springer, 2017.

\bibitem{BishopDFNNBZ2019}
Matt Bishop, Melissa Dark, Lynn Futcher, Johan Van~Niekerk, Ida Ngambeki, Somdutta Bose, and Minghua Zhu.
\newblock Learning principles and the secure programming clinic.
\newblock In Lynette Drevin and Marianthi Theocharidou, editors, {\em Information Security Education. Education in Proactive Information Security}, pages 16--29, Cham, 2019. Springer International Publishing.

\bibitem{SEICERTC}
{Carnegie Mellon University}.
\newblock {SEI CERT} {C} coding standard.
\newblock Accessed: 2024-02-15.

\bibitem{SEICERTC++}
{Carnegie Mellon University}.
\newblock {SEI CERT} {C}++ coding standard.
\newblock Accessed: 2024-02-15.

\bibitem{SEICERTJava}
{Carnegie Mellon University}.
\newblock {SEI CERT} {O}racle coding standard for {J}ava.
\newblock Accessed: 2024-02-15.

\bibitem{SEICERTTop10}
{Carnegie Mellon University}.
\newblock Top 10 secure coding practices.
\newblock Accessed: 2024-02-15.

\bibitem{Cenzic2009}
Cenzic.
\newblock Web application security trends report.
\newblock Technical report, Cenzic, http://www.cenzic.com/downloads/Cenzic\_AppSecTrends\_Q1-Q2-2009.pdf, Q1-Q2 2009.

\bibitem{ChiJB2013}
Hongmei Chi, Edward~L. Jones, and John Brown.
\newblock Teaching secure coding practices to {STEM} students.
\newblock In {\em Proceedings of the 2013 on InfoSecCD ’13: Information Security Curriculum Development Conference}, InfoSecCD ’13. ACM, October 2013.

\bibitem{CookC1979}
{Thomas D} Cook and {D T} Campbell.
\newblock {\em Quasi-Experimentation: Design and Analysis Issues for Field Settings}.
\newblock Houghton Mifflin, 1979.

\bibitem{DossYCAF2019}
Christopher Doss, Xiaohong Yuan, Varshar Chennakeshva, Aakiel Abernathy, and Kenneth Ford.
\newblock Modules for teaching secure in {A}ndroid application development.
\newblock In {\em Journal of The Colloquium for Information Systems Security Education}, volume~6, pages 13--13, 2019.

\bibitem{EarwoodYK2021}
Brandon Earwood, Jeong Yang, and Young~Rae Kim.
\newblock Effective learning of cybersecurity concepts with model-eliciting activities.
\newblock In {\em 2021 IEEE International Conference on Engineering, Technology \& Education (TALE)}, pages 01--07. IEEE, 2021.

\bibitem{Elva2022}
Rochelle Elva.
\newblock {TRAC}: An approach to teaching security-aware programming in undergraduate computer science courses.
\newblock In {\em Proceedings of the Computer 10th International Conference on Foundations of Computer Science \& Technology (FCST 2022), Zurich, Switzerland, May 21-22, 2022}, volume~12, 2022.

\bibitem{ENISA_threat_lanskape}
ENISA.
\newblock {ENISA} threat landscape 2023, 2023.
\newblock Accessed: 24-10-2023.

\bibitem{GhiglieriS2016}
Marco Ghiglieri and Martin Stopczynski.
\newblock Sec{L}ab: An innovative approach to learn and understand current security and privacy issues.
\newblock In Deborah Boisvert and Stephen~J. Zilora, editors, {\em Proceedings of the 17th Annual Conference on Information Technology Education and the 5th Annual Conference on Research in Information Technology, {SIGITE/RIIT} 2016, Boston, MA, USA, September 28 - October 1, 2016}, pages 67--72. {ACM}, 2016.

\bibitem{Haywood2013}
Adley Haywood, Huiming Yu, and Xiaohong Yuan.
\newblock Teaching {J}ava security to enhance cybersecurity education.
\newblock In {\em 2013 Proceedings of IEEE SoutheastCon}, IEEE SoutheastCon-Proceedings, pages 1--6. IEEE; UNF Student Branch, April 2013.

\bibitem{HowardL2003}
Michael Howard and David LeBlanc.
\newblock {\em Writing Secure Code}.
\newblock Microsoft Press, 2003.

\bibitem{HowardLV2005}
Michael Howard, David LeBlanc, and John Viega.
\newblock {\em 19 Deadly Sins of Software Security}.
\newblock McGraw-Hill Osborne Media, 2005.

\bibitem{KitchenhamC2007}
B~Kitchenham and S~Charters.
\newblock Guidelines for performing systematic literature reviews in software engineering.
\newblock Technical report, Technical report, EBSE Technical Report EBSE-2007-01, 2007.

\bibitem{MdunyelwaFN2019}
Vuyolwethu Mdunyelwa, Lynn Futcher, and Johan van Niekerk.
\newblock {\em An Educational Intervention for Teaching Secure Coding Practices}, pages 3--15.
\newblock Springer International Publishing, 2019.

\bibitem{CWE}
MITRE.
\newblock Common weakness enumeration.
\newblock Accessed: 2024-02-15.

\bibitem{NICE}
NICCS.
\newblock Workforce framework for cybersecurity ({NICE} framework).
\newblock Accessed: 2024-02-20.

\bibitem{OWASPSecCod}
OWASP.
\newblock {OWASP} secure coding practices-quick reference guide.
\newblock Accessed: 2024-02-20.

\bibitem{OWASPTop10}
OWASP.
\newblock {OWASP} top ten.
\newblock Accessed: 2024-02-15.

\bibitem{Pawelczak2020}
Dieter Pawelczak.
\newblock Teaching security in introductory {C}-programming courses.
\newblock In J~Domenech, P~Merello, E~DeLaPoza, and R~PenaOrtiz, editors, {\em HEAD'20: 6TH International Conference on Higher Education Advances}, pages 595--603. Inst Ciencias Educac; Univ Politecnica Valencia, Fac Administrac Direcc Empresas, Departamento Economia Ciencias Sociales; Centro Ingn Economica; European Union, European Social Fund, 2020.

\bibitem{PeltsvergerK2010}
Svetlana Peltsverger and Orlando Karam.
\newblock Is teaching with security in mind working?
\newblock In {\em 2010 Information Security Curriculum Development Conference}, InfoSecCD '10, page 15–20, New York, NY, USA, 2010. Association for Computing Machinery.

\bibitem{PetersenVK2015}
Kai Petersen, Sairam Vakkalanka, and Ludwik Kuzniarz.
\newblock Guidelines for conducting systematic mapping studies in software engineering: An update.
\newblock {\em Information and Software Technology}, 64:1--18, 2015.

\bibitem{PetticrewR2006}
Mark Petticrew and Helen Roberts.
\newblock {\em Systematic Reviews in the Social Sciences: A Practical Guide}.
\newblock Blackwell Publishing, 2006.

\bibitem{CWETop25}
{SANS Institute}.
\newblock {CWE} top 25 most dangerous software errors.
\newblock Accessed: 2024-02-15.

\bibitem{SharPSW2022}
Lwin~Khin Shar, Christopher~M Poskitt, Kyong~Jin Shim, and Li~Ying~Leonard Wong.
\newblock {XSS} for the masses: Integrating security in a web programming course using a security scanner.
\newblock In {\em Proceedings of the 27th ACM Conference on on Innovation and Technology in Computer Science Education Vol. 1}, pages 463--469, 2022.

\bibitem{StallingsB2007}
William Stallings and Lawrie Brown.
\newblock {\em Computer Security: Principles and Practice}.
\newblock Prentice Hall Press, USA, 1st edition, 2007.

\bibitem{TabassumWCL2018}
Madiha Tabassum, Stacey Watson, Bill Chu, and Heather~Richter Lipford.
\newblock Evaluating two methods for integrating secure programming education.
\newblock In {\em Proceedings of the 49th ACM Technical Symposium on Computer Science Education}, pages 390--395, 2018.

\bibitem{TangD2019}
Yuzhe Tang and Wenliang Du.
\newblock Hands-on labs for secure programming on modern trusted platforms.
\newblock In {\em The Colloquium For Information Systems Security Education {CISSE}}, jun 2019.

\bibitem{TaylorK2016}
Blair Taylor and Siddharth Kaza.
\newblock Security {Injections}@{Towson}: Integrating secure coding into introductory computer science courses.
\newblock {\em ACM Trans. Comput. Educ.}, 16(4), jun 2016.

\bibitem{TsipenyukCM2005}
Katrina Tsipenyuk, Brian Chess, and Gary McGraw.
\newblock Seven pernicious kingdoms: A taxonomy of software security errors.
\newblock {\em IEEE Security \& Privacy}, 3(6):81--84, 2005.

\bibitem{ViegaM2001}
John Viega and Gary~R McGraw.
\newblock {\em Building Secure Software: How to Avoid Security Problems the Right Way}.
\newblock Addison-Wesley, 2002.

\bibitem{SvabenskyVC2020}
Valdemar \v{S}v\'{a}bensk\'{y}, Jan Vykopal, and Pavel \v{C}eleda.
\newblock What are cybersecurity education papers about? {A} systematic literature review of {SIGCSE} and {ITiCSE} conferences.
\newblock In {\em Proceedings of the 51st ACM Technical Symposium on Computer Science Education}, SIGCSE '20, page 2–8, New York, NY, USA, 2020. Association for Computing Machinery.

\bibitem{WenK2019}
Shao{-}Fang Wen and Basel Katt.
\newblock Preliminary evaluation of an ontology-based contextualized learning system for software security.
\newblock In Shaukat Ali and Vahid Garousi, editors, {\em Proceedings of the Evaluation and Assessment on Software Engineering, {EASE} 2019, Copenhagen, Denmark, April 15-17, 2019}, pages 90--99. {ACM}, 2019.

\bibitem{WhitneyRCZ2015}
Michael Whitney, Heather Lipford-Richter, Bill Chu, and Jun Zhu.
\newblock Embedding secure coding instruction into the {IDE}: A field study in an advanced {CS} course.
\newblock In {\em Proceedings of the 46th ACM Technical Symposium on Computer Science Education}, pages 60--65, 2015.

\bibitem{WilliamsYYB2014}
Kenneth~A Williams, Xiaohong Yuan, Huiming Yu, and Kelvin Bryant.
\newblock Teaching secure coding for beginning programmers.
\newblock {\em Journal of Computing Sciences in Colleges}, 29(5):91--99, 2014.

\bibitem{WohlinRHORW2012}
Claes Wohlin, Per Runeson, Martin H{\"{o}}st, Magnus~C. Ohlsson, and Bj{\"{o}}rn Regnell.
\newblock {\em Experimentation in Software Engineering}.
\newblock Springer, 2012.

\bibitem{YuJBY2011}
Huiming Yu, Nadia Jones, Gina Bullock, and Xiaohong~Yuan Yuan.
\newblock Teaching secure software engineering: Writing secure code.
\newblock In {\em 2011 7th Central and Eastern European Software Engineering Conference (CEE-SECR)}, pages 1--5, Oct 2011.

\end{thebibliography}
